\documentclass[amsmath,amssymb,aps,pre,reprint, showpacs, superscriptaddress]{revtex4-1}
\usepackage{graphicx}
\usepackage{hyperref}

\begin{document} 
\title[]{Cycling tames power fluctuations near optimum efficiency}
\author{Viktor Holubec}
\email{viktor.holubec@gmail.com}
\affiliation{ 
Institut f{\"u}r Theoretische Physik, 
Universit{\"a}t Leipzig, 
Postfach 100 920, D-04009 Leipzig, Germany
}
\affiliation{ 
 Charles University,  
 Faculty of Mathematics and Physics, 
 Department of Macromolecular Physics, 
 V Hole{\v s}ovi{\v c}k{\' a}ch 2, 
 CZ-180~00~Praha, Czech Republic 
}
\author{Artem Ryabov}
\affiliation{ 
 Charles University,  
 Faculty of Mathematics and Physics, 
 Department of Macromolecular Physics, 
 V Hole{\v s}ovi{\v c}k{\' a}ch 2, 
 CZ-180~00~Praha, Czech Republic 
}
\date{\today} 
\begin{abstract} 
According to the laws of thermodynamics, no heat engine can beat the efficiency of a Carnot cycle.
This efficiency traditionally comes with vanishing power output and practical designs, optimized
for power, generally achieve far less. Recently, various strategies to obtain Carnot's efficiency at
large power were proposed. However, a thermodynamic uncertainty relation implies that steady-state heat engines can operate in this regime only at the cost of large fluctuations that render them
immensely unreliable. Here, we demonstrate that this unfortunate trade-off can be overcome by
designs operating cyclically under quasi-static conditions. The experimentally relevant yet exactly
solvable model of an overdamped Brownian heat engine is used to illustrate the formal result. Our
study highlights that work in cyclic heat engines and that in quasi-static ones are different stochastic processes.
\end{abstract}


\maketitle  
\section{Introduction}
Conversion of disordered energy (heat) into a directed motion (work) propels not only the industry but also the Nature itself through photosynthesis.
 According to the laws of thermodynamics, the efficiency $\eta = W/Q_h$ of this conversion is bounded from above by Carnot's efficiency $\eta_C = 1-T_c/T_h$~\cite{Callen2006}. 
The average heat $Q_h$ from a heat source can at most yield the average work $W=\left<w\right> = \eta_C Q_h$, remaining energy must be transferred into a heat sink. The upper bound is saturated if the temperatures of the hot and cold heat reservoirs assume constant values $T_h$ and $T_c$, respectively, and if the heat engine (HE) operates reversibly. Also, it is frequently argued that $\eta_C$ can be reached only if the engine operates on an infinite time scale $t_p$ with vanishing output power $P=W/t_p$. Recently, this claim has been seriously challenged \cite{Benenti2011,Allahverdyan2013,Campisi2016,Lee2016,Holubec2017a,
Polettini2017,Holubec2017,Pietzonka2018,Brandner2015,
Shiraishi2016,Shiraishi2017,Dechant2018,Solon2018,Shiraishi2018}. 

It was shown that either using a special coupling between subsystems \cite{Allahverdyan2013}, working substances close to criticality \cite{Campisi2016,Holubec2017a}, or scalings leading to vanishing system relaxation times \cite{Polettini2017,Holubec2017,Pietzonka2018}, it is possible to asymptotically reach $\eta_C$ with $P>0$. Although the HEs used for derivation of the last-mentioned results obey the trade-off bounds $P \le C(\eta_C - \eta)$ \cite{Brandner2015,Shiraishi2016,Dechant2018}, they can operate with $\eta=\eta_C$ and $P>0$ since the parameter $C$ generally diverges with vanishing system relaxation time~\footnote{For example, the constant $\chi = C/T_c \eta$ in the bound (97) in Ref.~\cite{Dechant2018} diverges for vanishing relaxation time of the momentum $1/\gamma$, and the bound derived in Ref.~\cite{Shiraishi2016} diverges for diverging transition rates \cite{Pietzonka2018,Polettini2017}}. 

However, it was suggested that the price one has to pay for overcoming the trade-off between power and efficiency are large power fluctuations \cite{Holubec2017a,Pietzonka2018}. In the critical heat engine \cite{Campisi2016}, the fluctuations almost surely dominate the averages \cite{Holubec2017a} and also steady state HEs (SSHEs) exhibit large power fluctuations \cite{Pietzonka2018}. 

Here, we show that such a trade-off does not exist for quasi-static cyclic HEs (CHEs) with controllable relaxation times. These machines can work with $\eta$ asymptotically close to $\eta_C$ at $P>0$ with vanishing fluctuations. Specifically, we show that both the work and power fluctuation $\tilde{\sigma}_P = \sigma_W/W = \sqrt{\left<w^2\right> - W^2}/W$ and the Fano factor for work $\sigma^2_W/W$ are finite and can even vanish.

Our results highlight that the work done by CHEs and the work done by SSHEs are two different stochastic processes. Although their mean values can be equal \cite{Raz2016,Rotskoff2017,Ray2017}, their fluctuations are qualitatively different. The work in the SSHEs obeys thermodynamic uncertainty relations~\cite{Barato2015,Pietzonka2016,Pietzonka2017a,
Gingrich2016,Horowitz2017} which imply that the Fano factor for the output work diverges if the efficiency reaches $\eta_C$ \cite{Pietzonka2018}. The work in the CHEs obeys no such relation and it is possible to construct a CHEs operating with Carnot's efficiency and delivering a persistent deterministic power output.



  
\section{Cyclic heat engines}

\begin{figure}[t!]
\centering
{\includegraphics[width=0.9\columnwidth]{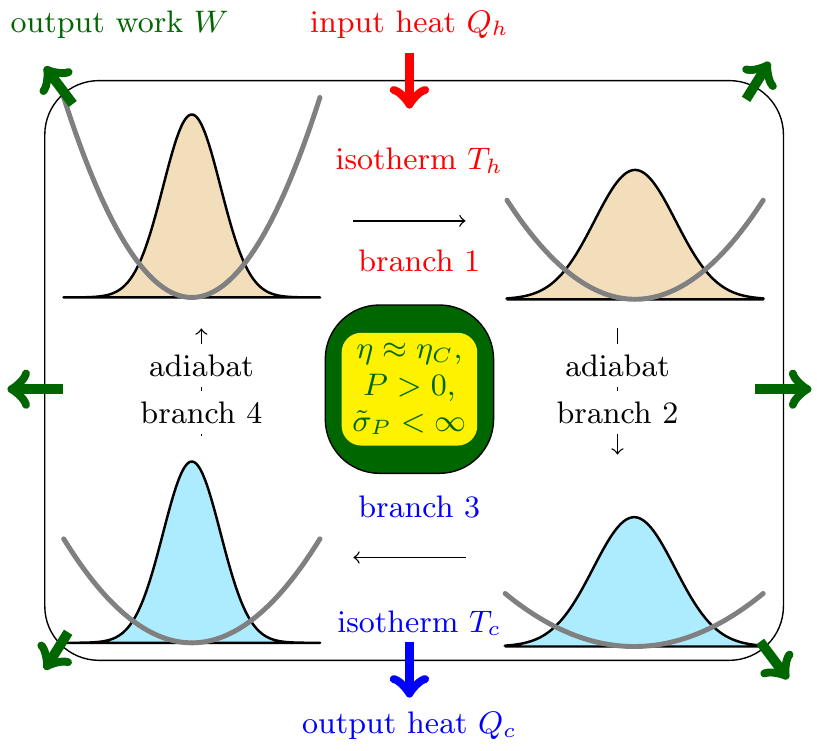}}
\caption{(Color online) The operational cycle of considered cyclic heat engines (CHEs). Gray lines depict the Hamiltonian \eqref{eq:arb_pot} and the shaded areas stand for the probability density of the particle position during the cycle.}
    \label{fig:fig1}
\end{figure}

Consider a periodically driven HE operating along a quasi-static Carnot cycle composed of two isotherms connected by two adiabats. For concreteness, we consider a one-dimensional system with the Hamiltonian
\begin{equation}
H(x,t) = k(t) x^{2n}/{2n},\quad n=1,2,\dots,
\label{eq:arb_pot}
\end{equation}
where $k=k(t)$ controls its stifness and $x=x(t)$ is a continuous stochastic process describing microstate of the system. The Hamiltonian~\eqref{eq:arb_pot} serves as a mere illustration. Our main results are valid for arbitrary thermodynamic systems which can operate quasi-statically, including many-dimensional systems with momentum degrees of freedom and systems with discrete state-space.

The operational cycle of the engine is depicted in Fig.~\ref{fig:fig1}. During the hot isotherm at $T_h$ (branch 1) and during the subsequent adiabat (branch 2), the Hamiltonian opens ($\dot{k}\ge 0$) and the system performs work $w_h = -\int_{1,2} dt \partial_t H(x,t) = -\int_{1,2} dt \dot{k}(t) x(t)^{2n}/{2n}$ on the environment (the integration runs over the branches 1 and 2). During the rest of the cycle, the Hamiltonian closes ($\dot{k} \le 0$) and the engine consumes work $- w_c = \int_{3,4} dt \partial_t H(x,t) = -\int_{3,4} dt \dot{k}(t) x(t)^{2n}/{2n}$. The heat on average enters the system during the hot isotherm and leaves it during the cold one (branch 3). We denote the duration of the $i$th branch as $t_i$ and as $t_p = t_1 + t_2 + t_3 + t_4$ the duration of the whole cycle. 

The average thermodynamics of the engine observed after averaging work and heat over many cycles 
is that of a standard reversible Carnot cycle. Namely, a combination of the first and the second law of thermodynamics implies that the average output work $W$ is given by \cite{Callen2006, SI}
\begin{equation}
W = \left< w \right> = \left< w_h+w_c \right> = Q_h - Q_c = (T_h - T_c)\Delta S,
\label{eq:power_Wirr}
\end{equation}
where $\Delta S$ is the change of the system entropy during the hot isotherm. On the other hand, the work fluctuations depend both on the details of the Hamiltonian and on the way how the adiabatic branches are realized. 

By definition, no heat flows into the system during adiabatic branches. This condition can be realized in two physically different ways. (i) One ensures that no heat \emph{at all} flows between the system and the bath by performing the adiabats very fast, or by disconnecting the system from the reservoir. During these adiabatic branches, the system evolves deterministically regardless of the dynamics of the baths. In general, reconnecting the bath and the system at the end of such adiabat brings the system far from equilibrium. To keep the cycle quasi-static, it is necessary to secure that the system state just before the reconnection is identical with the equilibrium state corresponding to the bath temperature and system Hamiltonian at the time of reconnection. (ii) One ensures that no heat is interchanged \emph{on average} only by carefully controlling the system connected to the reservoir with varying temperature \cite{Martinez2015,Arold2018}. Due to the coupling to the bath, the system evolves during such adiabats stochastically. 

We start with the traditional adiabatic branches (i) where no heat at all is exchanged leading to a deterministic evolution of the system during the adiabats. Then the work PDF $p(w)$ can be expressed as an average over the distributions for internal energy increases $\Delta H_2$ and $\Delta H_4$ along the adiabatic branches $2$ and $4$, respectively \cite{SI}:
\begin{equation}
p(w) = \left< \delta\left\{
w - \left[W - \widetilde{\Delta H}_{2} - \widetilde{\Delta H}_{4}   \right] \right\}\right>,
\label{eq:WPD}
\end{equation}
where $\widetilde{\Delta H}_{i} = \Delta H_i - \left< \Delta H_i \right>$, $i=2,4$. The PDF for $\Delta H_2$ and $\Delta H_4$ can be constructed from the Boltzmann distribution $\rho(x,\tau_i) = \exp\left[-H(x,\tau_i)/k_BT(\tau_i) \right]/Z(\tau_i)$ corresponding to the system Hamiltonian and bath temperature at times $\tau_i$, $i=1,\dots,4$ delimiting the adiabatic branches. Here $k_B$ denotes the Boltzmann constant and $Z$ is the partition function.

The PDF \eqref{eq:WPD} allows us to calculate all moments of work: $\left<w^n\right> = \int_{-\infty}^{\infty}dw w^n \rho(w)$. For the case of infinitely fast adiabatic branches ($t_2 \to 0$ and $t_4 \to 0$), the microstate of the system during the adiabatic branches does not change. Assuming that the particle is at a microstate $x$ at the beginning of the first adiabat and at a microstate $y$ at the beginning of the second one, the energy differences in Eq.~\eqref{eq:WPD} read $\Delta H_{2} = H(x, t_1 + t_2) - H(x,t_1)$ and $\Delta H_{4} = H(y, t_p) - H(y,t_p - t_4)$ and the average therein must be taken over the PDF $\rho(x,t_1)\rho(y,t_p)$. The work and power fluctuation evaluated for the Hamiltonian \eqref{eq:arb_pot} are then given by \cite{SI}
\begin{equation}
\tilde{\sigma}_w = \frac{\sigma_w}{W} = \frac{\sqrt{\left<w^2\right> - W^2}}{W}  = \frac{1}{\sqrt{n}}\frac{k_B}{\Delta S}.
\label{eq:rel_work_var_psead}
\end{equation}
The function $\tilde{\sigma}_w = \tilde{\sigma}_P$, which quantifies observability of the average work and power at the Carnot efficiency, is thus finite and decreases both with the exponent $n$ in the Hamiltonian \eqref{eq:arb_pot} and with the change of the system entropy during the hot isotherm $\Delta S$.

During the adiabatic branches (i) performed in a finite time with the disconnected heat bath, the system undergoes a non-trivial evolution determined by the Hamiltonian (through Hamiltonian equations for classical systems and Schr\"{o}dinger equation in quantum cases). To get an analytical result valid for arbitrary $H$, we use the approximation that microstates occupied by the system at the beginning of the adiabats are independent from those occupied at their ends. Then, the assumption that the system is in equilibrium both before the beginning and after the end of the adiabats allows us to calculate the work fluctuation along similar lines as in the previous case. The result is \cite{SI}
\begin{equation}
\tilde{\sigma}_w  =  \frac{1}{\sqrt{n}}\frac{k_B}{\Delta S}\frac{\sqrt{1+(1-\eta_C)^2}}{\eta_C} \ge \frac{1}{\sqrt{n}}\frac{k_B}{\Delta S}.
\label{eq:rel_work_var_ad}
\end{equation}
Compared to the work fluctuation \eqref{eq:rel_work_var_psead}, $\tilde{\sigma}_w$ now depends on the temperatures of the two baths via the Carnot efficiency $\eta_C$. The additional factor is always greater than one and thus Eq.~\eqref{eq:rel_work_var_psead} for the cycle with instantaneous adiabatic branches sets the lower bound on \eqref{eq:rel_work_var_ad}. 

The work fluctuations \eqref{eq:rel_work_var_psead} and \eqref{eq:rel_work_var_ad} are always nonzero. Their origin can be mapped to disconnecting the system from the baths during the adiabatic branches. According to its definition $w = - \int_0^{t_p} dt \partial_t{H}(x,t)$, the work is in CHEs done only if the Hamiltonian changes in time.  Along a quasi-static process, the reservoir causes many transitions in the system on the time-scale on which the external parameter corresponding to the work (for example the stiffness $k$ here, a piston position in thermodynamics) is varied. The time spent by arbitrary quasi-static trajectory $x(t)$ in a microstate $y$ within the time window $[t,t+dt]$ is determined by the Boltzmann distribution $\rho(y,t)$. The work $w$ done during a quasi-static process along each trajectory is hence given by the average work $W = - \int dx \int_0^{t_p} dt \partial_t{H}(x,t) \rho(x,t) = (T_h - T_c)\Delta S$ 
\cite{Speck2004,Hoppenau2013,Holubec2014a, SI}.

Quasi-static Carnot cycles with adiabatic branches (ii) where the system can interchange heat with the bath thus yield sharp work PDF 
\begin{equation}
p(w) = \delta(w - W)
\label{eq:EQWPDF}
\end{equation}
with vanishing variance $\sigma_w^2$ and fluctuation $\tilde{\sigma}_w$. Different from Eqs.~\eqref{eq:rel_work_var_psead} and \eqref{eq:rel_work_var_ad}, this result does not depend on the system Hamiltonian. As one consequence, the large power fluctuations found in the critical heat engine \cite{Campisi2016,Holubec2017a} can be avoided by utilizing this type of quasi-static adiabatic branches.

\section{Comparison with steady state heat engines}

Steady state HEs are connected to the hot and to the cold reservoir simultaneously and operate in a non-equilibirum steady state. They obey the current fluctuation relations~\cite{Barato2015,Pietzonka2016,Pietzonka2017a,
Gingrich2016,Horowitz2017} which can be used to derive the inequality for the relative work and power variance~\cite{Pietzonka2018}
\begin{equation}
\tilde{\sigma}_{w_t}^2 
\ge \frac{2k_B T_c }{W_t}\frac{\eta}{\eta_C - \eta} = \frac{2k_B}{\Delta S_{t}}.
\label{eq:SS_atEC}
\end{equation}
Here, $W_t$ and $\Delta S_{t}$ are the work and entropy generated during time window $[0,t]$. The formula \eqref{eq:SS_atEC} is valid in the long time limit $t\to \infty$, when the PDF for work attains the large deviation form.

The formula \eqref{eq:SS_atEC} implies that it is not possible to construct a SSHE working with Carnot's efficiency $\eta = \eta_C$, delivering work with a finite fluctuation $\tilde{\sigma}_{w_t}$ and operating reversibly with $\Delta S_t = 0$, at the same time. The SSHEs operating with $\eta_C$ must either dissipate ($\Delta S_t > 0$) or yield diverging work fluctuations ($\tilde{\sigma}_{w_t}\to\infty$). This observation is a HE analogy of the result obtained for Brownian clocks \cite{Barato2016}.

Another striking difference between the CHEs and the SSHEs is revealed if we rewrite our findings for CHEs in terms of the Fano factor for work $\sigma_w^2/W$, which equals to the ratio of constancy $\Delta_P = \sigma^2_P t$, $t\gg 1$ \cite{Pietzonka2018} to the output power $P=W/t$. The formula \eqref{eq:rel_work_var_psead} for a CHE operating with Carnot efficiency gives
\begin{equation}
\frac{\Delta_P}{P} = \frac{\sigma_w^2}{W} = \frac{1}{n}\frac{T_h \eta_C k_B^2}{\Delta S}
\end{equation}
and thus the Fano factor is in this case finite. Equation \eqref{eq:rel_work_var_ad} yields analogous results and the Fano factor corresponding to the work PDF \eqref{eq:EQWPDF} even vanishes. 

On the other hand, Eq.~\eqref{eq:SS_atEC} for the SSHEs leads to
\begin{equation}
\frac{\Delta_{P_t}}{P_t} = \frac{\sigma_{w_t}^2}{W_t} \ge 2 k_B T_c\frac{\eta}{\eta_C-\eta}
\end{equation}
which diverges whenever $\eta\to \eta_C$. The work and power fluctuations in the CHEs and in the SSHEs operating with $\eta_C$ thus significantly differ.

One may object that these conclusions are based on a comparison of incompatible quantities -- variables measured per cycle for CHEs and variables measured over a long time for SSHEs.
Nevertheless, measuring the quantities for the CHEs over many cycles or over many independent systems does not alter the main conclusions. 
More precisely, averaging over $N$ independent CHEs or, equivalently, over $N$ cycles of a single CHEs, both the average output work $W$ and its variance $\sigma^2_w$ scale as $N$. Therefore, although the fluctuation $\tilde{\sigma}_w$ scales as $1/\sqrt{N}$, the ratio $\Delta_P /P = \sigma_w^2/W$ remains constant. 

The difference between work in CHEs and SSHEs lies in the very definitions of these variables. Work in CHEs is done only when an external parameter changes and under quasi-static conditions it is independent of the initial microstate of the system \cite{SI}. On the contrary, work in SSHEs is usually done when the microstate $x$ of the system changes. 
During this thermally-induced transition, the system internal energy is increased in ratchets \citep{Lee2016}, particles are transferred against gradients of chemical potential in thermochemical heat engines \citep{Polettini2017,Pietzonka2018}, etc. Such defined work depends on the initial and final points of the stochastic trajectory $\{x(t)\}_{t=0}^{t_p}$, which for example determine the increase in the internal energy in a ratchet, and thus it always fluctuates. Work in SSHEs hence lacks the self-averaging property of the work done in CHEs. It is rather similar to the heat $Q = \int_0^{t_p} dt \partial_x H(x,t) \dot{x}$ in CHEs which is interchanged with the bath also only if the system microstate changes.


Our analysis implies that the work done in SSHEs and that in CHEs represent two different stochastic processes which cannot be directly mapped onto each other. Nevertheless, such a mapping might be constructed if the different definitions of work in the two classes of HEs would be taken into account.

\section{Cyclic Brownian heat engine}

Let us now propose an actual CHE operating close to Carnot's efficiency while delivering a stable power output. Its engineering is rather straightforward, it can be performed with an arbitrary thermodynamic system capable of quasi-static operation. In order to further demonstrate that such a HE can operate in finite time, delivering a nonzero output power, we need a system with controllable relaxation time. A paradigmatic example of such a system from the field of stochastic thermodynamics \cite{sek10,sei12} is the overdamped Brownian HE~\cite{Schmiedl2008,Holubec2017,Martinez2017}.

The HE is based on an overdamped Browninan particle diffusing in a harmonic potential~\footnote{We leave aside the discussion whether the Hamiltonian $H(x,t) = k(t)x^2/2$, or rather the full Hamiltonian $H(x,t) = k(t)x^2/2 + p^2/2m$ including the particle mass $m$ and momentum $p$ should be used for describing thermodynamics of the overdamped particle \cite{Martinez2015,Arold2018}. It is not important for our demonstrative purposes and in both cases one can construct a Brownian HE operating close to $\eta_C$ at $P>0$ with finite fluctuation.} 
$U(x,t) = H(x,t) = k(t)x^2/2$,
whose dynamics obeys the Langevin equation
\begin{equation}
\dot{x} = - k x/\gamma + \sqrt{2 k_B T/\gamma} \zeta.
\label{eq:LE}
\end{equation}
Here, $\zeta$ is the Gaussian white noise with $\left<\zeta\right>=0$ and $\left<\zeta(t)\zeta(t')\right> = \delta(t-t')$. The relaxation time for the position, $\tau_x = \gamma/k$, can be easily controlled in experiments through the trap stiffness $k$. The friction coefficient $\gamma$ is assumed to be independent of $k$ (yet it may depend on the temperature). The model is valid if this relaxation time is much longer that the relaxation time for the momentum, $\tau_p = m/\gamma$, given by the ratio of the particle mass $m$ to the friction $\gamma$. The model~\eqref{eq:LE} is exactly solvable and it has been thoroughly investigated both theoretically \cite{Schmiedl2008,Holubec2017} and experimentally, using optical tweezers for generation of the potential \cite{Blickle2012,Martinez2016,Martinez2017}.

To demonstrate our results for instantaneous adiabatic branches (i), we periodically modulate the bath temperature $T$ and the trap stiffness $k$ using the Carnot-like driving depicted in Fig.~\ref{fig:fig1} with infinitely fast adiabatic branches. If the cycle is performed in a finite time $t_p$, with a non-vanishing relaxation time $\tau_x$, the system is during the cycle inevitably out of equilibrium and the HE efficiency is smaller than $\eta_C$. In order to realize the quasi-static Carnot cycle using a finite $t_p$, we thus need to use a very stiff trap, which makes $\tau_x \ll t_p$.

\begin{figure}[t!]
\centering
{\includegraphics[width=1.0\columnwidth]{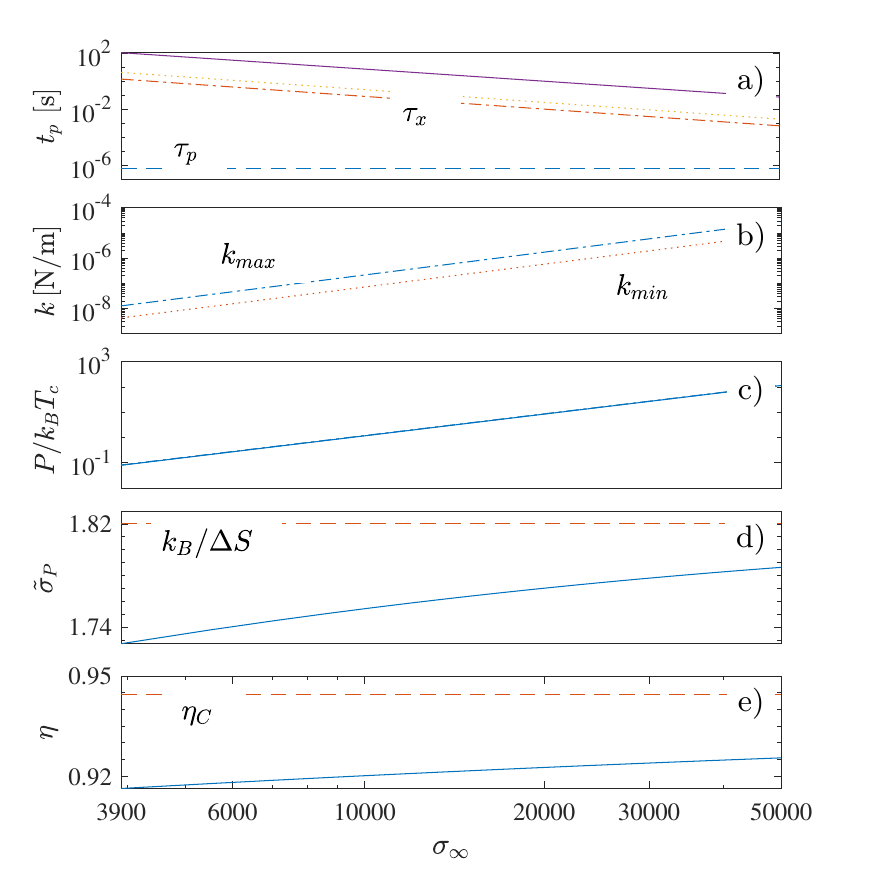}}
\caption{(Color online) Behavior of the overdamped Brownian HE with the scaling parameter $\sigma_\infty$. Time-scale separation between the cycle duration $t_p$ and the relaxation times $\tau_x$ (we show its smallest and largest value during the cycle) and $\tau_p$ is depicted in the panel a). In panel b), $k_{max}$ ($k_{min}$) stands for the maximum/minimum value of the stiffness during the cycle. The shown values of cycle durations $t_p$ and trap stiffnesses $k$ are reasonable from experimental perspective. In panels c) and d) we demonstrate divergence of output power $P$ and convergence of the relative power fluctuation $\tilde{\sigma}_P$ to $k_B/\Delta S$ as the efficiency $\eta$, shown in panel c), converges to $\eta_C$ for large values of $\sigma_\infty$.
   }
    \label{fig:fig2}
\end{figure}

In Figs.~\ref{fig:fig2} a) and b), we introduce a suitable scaling of the cycle duration $t_p$ and minimum and maximum trap stiffness $k$ during the cycle which, in the limit of infinite scaling parameter $\sigma_\infty$, leads to a HE operating with Carnot's efficiency and delivering an infinite power with fluctuation given by Eq.~\eqref{eq:rel_work_var_psead} with $n = 1$. The convergence of the output power, the power fluctuation and the efficiency to these values as the cycle becomes gradually quasi-static with increasing $\sigma_\infty$ is plotted in Figs.~c), d) and e), respectively.
The curves are plotted using experimentally motivated values of the model parameters \cite{Martinez2017}. Further details are given in Supplementary Material~\cite{SI}. The rest of our results can be tested along similar lines.

\section{Concluding remarks}

Unlike steady state heat engines (SSHEs), cyclic heat engines (CHEs) can theoretically operate reversibly with Carnot's efficiency $\eta_C$, delivering a large and stable power output $P$ with finite fluctuation and Fano factor. The main difference between the two classes of heat engines lies in the definitions of work in the two models. While the transitions caused in the system due to the contact with the bath lead to averaging of work in CHEs, such an averaging is not available for SSHEs. In the latter case, the work always depends on initial and final point of a trajectory and thus inevitably fluctuates. The recently proposed one-to-one mappings between SSHEs and CHEs \cite{Raz2016,Rotskoff2017,Ray2017} thus break down on the level of work fluctuations.


In practice, the described strategy does not allow to realize the strict limit $\eta=\eta_C$ at $P>0$ without breaking the system-reservoir time-scale separation used in standard thermodynamic models \cite{Shiraishi2017}. But it is possible to find parameter regimes where realizable systems operate with efficiencies close to $\eta_C$ and deliver large power $P$ with small fluctuation. Experimental realizations of such HEs are possible using current micro-manipulation techniques such as optical tweezers \cite{Blickle2012,Martinez2016}. Finally we stress that our results are valid for general HEs, including intensively studied quantum models \cite{Kosloff2017}.

\begin{acknowledgments}
We thank P. Pietzonka and A. Dechant for stimulating correspondence and to K. Kroy for helpful discussions. We gratefully acknowledges financial support by the Czech Science Foundation (project No. 17-06716S). VH also thanks for support by the Humboldt foundation.
\end{acknowledgments}

\bibliographystyle{apsrev4-1}	
\bibliography{references}


\end{document}